# A Cost Minimization Approach to Synthesis of Linear Reversible Circuits

Ben Schaeffer, Marek Perkowski

Electrical and Computer Engineering Department, Portland State University, Portland, Oregon, USA

{bms, mperkows}@cecs.pdx.edu

**Abstract.** This paper presents a heuristic cost minimization approach to synthesizing linear reversible circuits. Two bidirectional linear reversible circuit synthesis methods are introduced, the Alternating Elimination with Cost Minimization method (AECM) and the Multiple CNOT Gate method (MCG). Algorithms, example syntheses, and extensions to these methods are presented. An MCG variant which incorporates line reordering is introduced. Tests comparing the new cost minimization methods with the best known method for large circuits are presented. Results show that of the three methods MCG had the lowest average CNOT gate counts for linear reversible circuits up to 24 lines, and that AECM had the lowest counts between 28 and 60 lines.

## 1 Introduction

Linear reversible circuits, which are circuits which employ only controlled-NOT (CNOT) gates, play a fundamental role in both reversible and quantum computing. The most basic form of linear reversible circuit synthesis is built on a $GF(2)$-based variant of Gaussian Elimination which uses an invertible $n \times n$ Boolean matrix $M$ as its input and produces an $n \times n$ linear reversible circuit as its output [1,2]. Using $GF(2)$ input vector $x$ and output vector $y$ this circuit performs the function $y = Mx$. Synthesis using Gaussian Elimination produces circuits with $O(n^2)$ CNOT gates in $O(n^3)$ time. In [1] Patel et. al. introduced "Algorithm 1" which adapted the Four Russians Method for $GF(2)$ matrix inversion [3,4] to the case of linear reversible circuit synthesis. By using a strategy of processing two or more matrix columns simultaneously, "Algorithm 1" produces circuits with $O(n^2/\log_2 n)$ CNOT gates in $O(n^3/\log_2 n)$ time. Since the introduction of "Algorithm 1" little else has been published about improving the gate count in linear reversible circuit synthesis. While "Algorithm 1" is "asymptotically optimal up to a multiplicative constant" [1] this algorithm is too simplistic to find an exact minimum solution to the function $y = [x_1, x_1 \oplus x_2, x_1 \oplus x_2 \oplus x_3, x_1 \oplus x_2 \oplus x_3 \oplus x_4]^T$ illustrated in Fig. 1 [2].

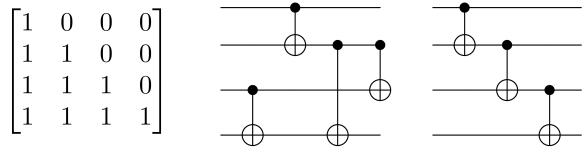

**Fig. 1.** A 4×4 linear function synthesized with "Algorithm 1" (middle) and an exact minimum circuit realization (right).

As an alternative to "Algorithm 1", two bidirectional linear reversible circuit synthesis methods – the Alternating Elimination with Cost Minimization method (AECM) and the Multiple CNOT Gate method (MCG) – are introduced herein. These methods were developed to more efficiently synthesize circuits of up to 64 lines (a.k.a. wires) which do not use ancilla lines. The efficiency comparison of AECM, MCG and "Algorithm 1" is shown in the Section 5.

AECM is built on the Alternating Elimination method [2] while MCG is based solely on cost minimization and is outside the Gaussian Elimination family of methods. Alternating Elimination extends the Gaussian Elimination approach of forward substitution and backward elimination to process diagonal matrix elements iteratively, in each iteration solving a diagonally intersecting row and column. Since there are $n!$ possible orderings of diagonal matrix elements, Alternating Elimination can generate a large number of functionally equivalent circuit solutions which have a range of CNOT gate counts.

Both AECM and MCG lend themselves to parallelization and can be extended to perform deeper, albeit slower, syntheses. In comparison with "Algorithm 1", which sequentially solves columns before rows, AECM and MCG solve rows and columns in a data dependent, nondeterministic order which provides the greatest cost reduction. Both AECM and MCG determine cost by means of a heuristic function which depends on a $GF(2)$ linear function's matrix and its inverse. When synthesis begins these matrices correspond to the input function specification, and as CNOT gates are synthesized these matrices correspond to a remainder function more closely resembling the identity matrix. The main heuristic cost function is defined as

$$\sum_{i=1}^{n}\sum_{j=1}^{n}\left(M_{(i,j)} \oplus I_{(i,j)}\right) + \left(M^{-1}_{(i,j)} \oplus I_{(i,j)}\right) \quad (1)$$

As seen in the above formula, the cost function is a sum of two components. The first is the number of differences between a linear function's matrix and the identity matrix. The second is the number of differences between the inverse of that linear function and the identity matrix. An alternative cost function

$$\sum_{i=1}^{n}\sum_{j=1}^{n} M_{(i,j)} + M^{-1}_{(i,j)} - 2n \quad (2)$$

is based on sparseness of a matrix and its inverse. Using this alternative cost function facilitates synthesis of a permutation of an input linear reversible function specification which will be discussed in Section 3.

The organization of this paper is as follows. Section 2 introduces and discusses algorithms for AECM and MCG. Section 3 illustrates algorithm flow of a nonconvergent MCG synthesis of a 5×5 linear function; next it compares AECM, MCG, and "Algorithm 1" synthesis of a 6×6 linear function; and lastly it illustrates line-reordering MCG synthesis. Section 4 shows numerical results of comparative testing of the different synthesis methods. Section 5 briefly discusses additional strategies to improve AECM and MCG. Section 6 concludes the paper.

## 2   Algorithms and Discussion

### 2.1   Algorithms

The algorithms for *Diagonalize()*, *AECM()*, *MCG()*, *Cost()*, and *ApplyCNOT()* are shown below in pseudocode form. A discussion of the algorithms follows.

```
Diagonalize(Matrix m, Matrix mi, integer threshold,
  CNOTgate list inputgates, CNOTgate list outputgates,
  integer diagonal)
c1 := Cost(m) + Cost(mi)
lastdiagonal := m.columns - 1
FOR i FROM 0 TO lastdiagonal
  IF NOT(i = diagonal)
    tempgate.control := i
    tempgate.target := diagonal
    tempgate.type := output
    improvement := ImpfromCNOT(m, mi, tempgate)
    IF (improvement >= 2)
      outputgates.Add(tempgate)
      ApplyCNOT(tempgate, m, mi)
      c1 := c1 - improvement
      IF (c1 <= threshold) return
    tempgate.control := diagonal
    tempgate.target := i
    tempgate.type := input
    improvement := ImpfromCNOT(m, mi, tempgate)
    IF (improvement >= 2)
      inputgates.Add(tempgate)
      ApplyCNOT(tempgate, m, mi)
      c1 := c1 - improvement
      IF (c1 <= threshold) return
  IF (m[diagonal][diagonal] = 0)
    improvementmax := -MAXLINES * MAXLINES * 2
    FOR i FROM 0 TO lastdiagonal
      IF (NOT(i = diagonal))
        IF (m[i][diagonal] = 1)
```

```
          tempgate.control := i
          tempgate.target := diagonal
          tempgate.type := output
          improvement := ImpfromCNOT(m, mi, tempgate)
          IF (improvement > improvementmax)
            improvementmax := improvement
            bestCNOT := tempgate
        IF (m[diagonal][i] = 1)
          tempgate.control := diagonal
          tempgate.target := i
          tempgate.type := input
          improvement := ImpfromCNOT(m, mi, tempgate)
          IF (improvement > improvementmax)
            improvementmax := improvement
            bestCNOT := tempgate
    IF (bestCNOT.type = output)
      IF (IsRedundant(bestCNOT, outputgates))
        outputgates.RemoveLast(bestCNOT)
      ELSE
        outputgates.Add(bestCNOT)
    ELSE
      IF (Redundant(bestCNOT, inputgates))
        inputgates.RemoveLast(bestCNOT)
      ELSE
        inputgates.Add(bestCNOT)
    ApplyCNOT(bestCNOT, m, mi)
    c1 := c1 - improvementmax
    IF (c1 <= threshold) return
  FOR i FROM 0 TO lastdiagonal
    IF (m[i][diagonal] = 1 AND NOT(i = diagonal))
      bestCNOT.control := diagonal
      bestCNOT.target := i
      bestCNOT.type := output
      improvementmax := ImpfromCNOT(m, mi, bestCNOT)
      tempgate.target := i
      FOR j FROM i + 1 TO lastdiagonal
        IF (m[j][diagonal] = 1 AND NOT(j = diagonal))
          tempgate.control := j
          improvement := ImpfromCNOT(m, mi, tempgate)
          IF (improvement > improvementmax)
            bestCNOT := tempgate
            improvementmax := improvement
      outputgates.Add(bestCNOT)
      ApplyCNOT(bestCNOT, m, mi)
      c1 := c1 - improvementmax
```

```
      IF (c1 <= threshold) return
    FOR i FROM 0 TO lastdiagonal
      IF (m[diagonal][i] = 1 AND NOT(i = diagonal))
        bestCNOT.control := i
        bestCNOT.target := diagonal
        bestCNOT.type := input
        improvementmax := ImpfromCNOT(m, mi, bestCNOT)
        tempgate.control := i
        FOR j FROM i + 1 TO lastdiagonal
          IF (m[diagonal][j] = 1 AND NOT(j = diagonal))
            tempgate.target := j
            improvement := ImpfromCNOT(m, mi, tempgate)
            IF (improvement > improvementmax)
              bestCNOT := tempgate
              improvementmax := improvement
        //commit to best gate
        inputgates.Add(bestCNOT)
        ApplyCNOT(bestCNOT, m, mi)
        c1 := c1 - improvementmax
        IF (c1 <= threshold) return

  AECM(Matrix m, Matrix mi, integer threshold,
    CNOTgate list inputgates, CNOTgate list outputgates)
    c1 := Cost(m) + Cost(mi)
    WHILE (c1 > threshold)
      bestcandidatefound := false
      FOR i := 0 TO m.columns - 1
        IF (bestcandidatefound) THEN
          IF (NOT DiagonalSolved(m, i))
            m3 := m
            mi3 := mi
            inputgates3 := inputgates
            outputgates3 := outputgates
            Diagonalize(m3, mi3, threshold,
              inputgates3, outputgates3, i)
            c3 := Cost(m3) + Cost(mi3)
            gates3 := m3.size + mi3.size - m.size
              - mi.size
            gain3 := (c1 - c3)/gates3
            IF (gain3 > gain2) THEN
              m2 := m3
              mi2 := mi3
              c2 := c3
              gates2 := gates3
              inputgates2 := inputgates3
```

```
              outputgates2 := outputgates3
        ELSE
          IF (NOT DiagonalSolved(rem, i)) THEN
            bestcandidatefound := true
            m2 := m
            mi2 := mi
            inputgates2 := inputgates
            outputgates2 := outputgates
            Diagonalize(m2, mi2, threshold,
              inputgates2, outputgates2, i)
            c2 := Cost(m2) + Cost(mi2)
            gates2 := m2.size + mi2.size - m.size
              - mi.size
            gain2 := (c1 - c2)/gates2
    m := m2
    mi := mi2
    inputgates := inputgates2
    outputgates := outputgates2
    c1 := c2
  IF (threshold = 0) THEN
    append outputgates in reverse order TO inputgates
  return c1

MCG(Matrix m, CNOTgate list inputgates,
  CNOTgate list outputgates)
  convergent := true
  mi := GF2MatrixInverse(m)
  c1 := Cost(m) + Cost(mi)
  c2 := c1
  WHILE (c1 > 0)
    FOR each gate1 in list allinputandoutputCNOTgates
      ApplyCNOT(gate1, m, mi)
      IF Cost(m) = 0 THEN
        inputgates.Add(gate1)
        exit WHILE
      FOR each gate2 in list allinputandoutputCNOTgates
        IF gate1 = gate2 THEN
          continue for
        ApplyCNOT(gate2, m, mi)
        c3 := Cost(m) + Cost(mi)
        IF (c2 > c3) THEN
          c2 := c3
          bestgate1 := gate1
          bestgate2 := gate2
        ApplyCNOT(gate2, m, mi) //undo operations
```

```
          ApplyCNOT(gate1, m, mi)
      IF (c2 = c1) THEN
        convergent := false
        c1 := AECM(m, mi, c1 - 1, inputgates, outputgates)
        c2 := c1
      ELSE
        c1 := c2
        IF (bestgate1.type = output)
          outputgates.Add(bestgate1)
        ELSE
          inputgates.Add(bestgate1)
        IF (bestgate2.type = output)
          outputgates.Add(bestgate2)
        ELSE
          inputgates.Add(bestgate2)
        ApplyCNOT(bestgate1, m, mi)
        ApplyCNOT(bestgate2, m, mi)
  AppendOutputgatesInReverseOrderToInputgates()
  return convergent

Cost(Matrix m)
  c1 := 0
  FOR row FROM 0 TO m.rows - 1
    FOR column FROM 0 TO m.columns - 1
      IF m[row][column] !:= I[row][column] THEN
        c1 := c1 + 1
  return c1

ApplyCNOT(CNOT g, Matrix m, Matrix mi)
  IF (g.type = output)
    FOR row i 0 TO m.rows - 1 //note rows := columns
      m[g.target][i] :=
        m[g.target][i] EXOR m[g.control][i]
      mi[i][g.control] :=
        mi[i][g.control] EXOR mi[i][g.target]
  ELSE
    FOR row i 0 TO m.rows - 1
      m[i][g.control] :=
        m[i][g.control] EXOR m[i][g.target]
      mi[g.target][i] :=
        mi[g.target][i] EXOR mi[g.control][i]
```

## 2.2 Discussion

Because AECM is based on Alternating Elimination, it will always converge to a solution for any linear function matrix given as input [2]. The AECM method iteratively compares $O(n)$ matrix diagonalizations and then commits to the diagonalization which produces the greatest cost reduction per CNOT gate ratio. In comparison with MCG, which requires the cost of the remainder function to be lower with each iteration, AECM can commit to using CNOT gate sequences which causes the cost to increase and therefore cannot be trapped in a local minimum.

The AECM diagonalization function has four stages, and in each stage the changes in cost of choosing candidate CNOT gates are compared. The first stage performs preprocessing through an $O(n)$ search to find row and column forward substitutions which lower the cost by at least two. Using a cost reduction of at least two is based on testing which showed that in over half the syntheses examined it produced lower CNOT gate counts than using a cost reduction of at least one or skipping the preprocessing stage. Using a cost reduction of at least three was in some instances superior and in other instances inferior to using at least two. Each CNOT gate synthesized in this stage replaces two or more CNOT gates which would have been synthesized in the third and fourth stages.

If the diagonal matrix element associated with the current iteration is a 0, then a second stage is used which performs forward substitution. In this second stage either a row or column forward substitution is chosen through an $O(n)$ search to find the CNOT gate which establishes a 1 on the diagonal and results in the lowest cost remainder function. When it is necessary to perform a forward substitution, a check is made to ensure that the forward substitution CNOT gate was not synthesized in the first stage. This situation is unusual but possible. In these cases the CNOT gate list can be rearranged to detect pairs of identical CNOT gates. Because CNOT gates are self-inverse [2], all detected identical CNOT gate pairs can be erased.

In the third stage $O(n)$ row-based backward eliminations are performed to process column elements which are equal to 1. Unlike Gauss-Jordan Elimination which performs eliminations using the diagonally intersecting row, here each row elimination employs an $O(n)$ search to find the lowest cost backward elimination operation. Similarly in the fourth stage $O(n)$ column-based backward eliminations are performed to process row elements which are equal to 1, each employing an $O(n)$ search to find the lowest cost backward elimination operation.

Performing one row or column addition with a cost difference computation takes $O(n)$ time. Therefore the entire AECM diagonalization function takes $O(n) \cdot (O(n) + O(n) + O(n^2) + O(n^2)) \approx O(n^3)$ time. Since the outer AECM loop requires $O(n)$ iterations through $O(n)$ comparisons, the total time is $O(n) \cdot O(n) \cdot O(n^3) \approx O(n^5)$.

In order to support partial syntheses the AECM algorithm uses the parameter *threshold*. Using AECM with *threshold* = 0 causes a complete synthesis to be performed. Using AECM with larger *threshold* values causes synthesis to terminate when the cost of the remainder function *c1* goes below *threshold*. In the algorithm's outermost loop, CNOT gate selection is performed by comparing *gain3* with *gain2*. These gain values are computed as $(Cost(F_{k-1}) - Cost(F_k))/(Gates(F_k) - Gates(F_{k-1}))$ for remainder function $F$ at iteration $k$. The AECM algorithm can be extended to handle occurrences in which these ratios are equal, thus facilitating algorithm extensions such as recursion and probabilistic gate selection.

The MCG synthesis method performs synthesis with linear functions composed of two CNOT gates, but in general this approach can be extended to three or more CNOT gates at the expense of increased computation time. The two-CNOT-gate functions can be categorized as one of three types: 1) functions of two elementary row operations corresponding to two CNOT gates synthesized from output towards input; 2) functions of two elementary column operations corresponding to two CNOT gates synthesized from input towards output; 3) functions of one elementary row operation and one elementary column operation representing one CNOT gate synthesized from output towards input and another synthesized from input towards output. The MCG method iteratively compares the cost of applying all possible two-CNOT-gate functions and commits to the pair of CNOT gates which produces the greatest reduction in cost. In the event that the cost reaches a local minimum, synthesis temporarily switches to AECM until cost drops below the local minimum cost. In this situation a flag is set indicating that MCG failed to converge and MCG synthesis resumes. In each iteration MCG retrieves two-CNOT-gate functions from an $O(n^4)$ length list. Performing elementary row or column operations and cost difference computations on each two-CNOT-gate function requires $O(n)$ time. Since the maximum cost is $2n^2$, the smallest cost 0, and the minimum cost reduction is 1 at each iteration, the two-CNOT-gate function search takes at most $O(n^2)$ outermost loop iterations. Therefore the total time is $O(n^4) \cdot O(n) \cdot O(n^2) \approx O(n^7)$.

Like AECM, MCG can be extended to perform more sophisticated gate selections in iterations where multiple minimum-cost alternatives exist. This will be demonstrated later using a probabilistic gate selection. Also, MCG's speed can be improved by using precalculated two-CNOT-gate functions. In the above MCG algorithm all possible CNOT gate sequences are generated, and many will be redundant. For instance, the two-CNOT-gate function CNOT(1, 2) followed by CNOT(3, 4) is equivalent to CNOT(3, 4) followed by CNOT(1, 2). If MCG is extended to use three-CNOT-gate functions, a greater variety of redundant sequences will be generated.

## 3     Example Syntheses

$$\begin{bmatrix} 1 & 0 & 0 & 1 & 1 \\ 0 & 1 & 1 & 0 & 1 \\ 0 & 1 & 1 & 1 & 0 \\ 1 & 0 & 1 & 1 & 0 \\ 1 & 1 & 0 & 0 & 1 \end{bmatrix}, \begin{bmatrix} 1 & 1 & 1 & 0 & 0 \\ 1 & 1 & 0 & 1 & 0 \\ 1 & 0 & 1 & 0 & 1 \\ 0 & 1 & 0 & 1 & 1 \\ 0 & 0 & 1 & 1 & 1 \end{bmatrix}$$

**Fig. 2.** Example of a *GF*(2) linear function, represented as a 5×5 matrix and its inverse, for which MCG fails to converge.

$$\begin{bmatrix} 1 & 0 & 0 & 0 & 0 \\ 0 & 1 & 1 & 0 & 1 \\ 0 & 1 & 1 & 1 & 0 \\ 0 & 0 & 1 & 0 & 1 \\ 0 & 1 & 0 & 1 & 0 \end{bmatrix}, \begin{bmatrix} 1 & 0 & 0 & 0 & 0 \\ 0 & 1 & 0 & 1 & 0 \\ 0 & 0 & 1 & 0 & 1 \\ 0 & 1 & 0 & 1 & 1 \\ 0 & 0 & 1 & 1 & 1 \end{bmatrix}$$

**Fig. 3.** Remainder linear function and its inverse after MCG employs AECM to produce four CNOT gates from the linear function and inverse from Fig. 2.

$$\begin{bmatrix} 1 & 0 & 0 & 0 & 0 \\ 0 & 1 & 0 & 0 & 0 \\ 0 & 0 & 1 & 1 & 0 \\ 0 & 0 & 1 & 0 & 1 \\ 0 & 0 & 0 & 1 & 0 \end{bmatrix}, \begin{bmatrix} 1 & 0 & 0 & 0 & 0 \\ 0 & 1 & 0 & 0 & 0 \\ 0 & 0 & 1 & 0 & 1 \\ 0 & 0 & 0 & 0 & 1 \\ 0 & 0 & 1 & 1 & 1 \end{bmatrix}$$

**Fig. 4.** Remainder linear function and its inverse after MCG produces two CNOT gates from the linear function and inverse from Fig. 3.

$$\begin{bmatrix} 1 & 0 & 0 & 0 & 0 \\ 0 & 1 & 0 & 0 & 0 \\ 0 & 0 & 1 & 1 & 0 \\ 0 & 0 & 0 & 1 & 1 \\ 0 & 0 & 0 & 0 & 1 \end{bmatrix}, \begin{bmatrix} 1 & 0 & 0 & 0 & 0 \\ 0 & 1 & 0 & 0 & 0 \\ 0 & 0 & 1 & 1 & 1 \\ 0 & 0 & 0 & 1 & 1 \\ 0 & 0 & 0 & 0 & 1 \end{bmatrix}$$

**Fig. 5.** Remainder linear function and its inverse after MCG produces two CNOT gates from the linear function and inverse from Fig. 4.

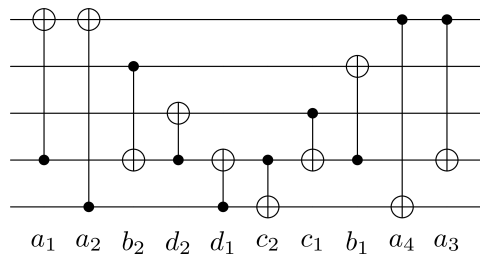

$a_1 \quad a_2 \quad b_2 \quad d_2 \quad d_1 \quad c_2 \quad c_1 \quad b_1 \quad a_4 \quad a_3$

**Fig. 6.** MCG circuit realization from the linear functions and inverses in Fig. 2 through Fig. 5.

The following summarizes MCG synthesis of the linear function in Fig. 2 to produce the linear reversible circuit in Fig. 6. Initially the convergence flag is set to true and cost is computed to be 20. In the first iteration the search for a two-CNOT-gate function which would lower the cost to 19 or less fails. The convergence flag is set to

*false* and AECM is called to perform a partial synthesis with *threshold* = 19. AECM selects the first row and column to be diagonalized and it synthesizes four CNOT gates, $a_1$ through $a_4$, resulting in the remainder function shown in Fig. 3. The cost of the remainder function and its inverse is 16. In the second iteration gates $b_1$ and $b_2$ are found to reduce the cost to 11, resulting in the second remainder function shown in Fig. 4. In the third iteration gates $c_1$ and $c_2$ are found to reduce the cost to 5, resulting in the third remainder function shown in Fig. 5. In the fourth iteration gates $d_1$ and $d_2$ are found to reduce the cost to 0, resulting in both the remainder function and its inverse becoming equal to the identity matrix indicating synthesis is complete. The final CNOT gate count of the circuit in Fig. 6 is 10, which is one gate above the exact minimum. The convergence flag plays no role in synthesis but was created to be used in statistics that correlate the increase in total CNOT gate count with nonconvergence.

The linear function in Fig. 2 is unusual as tests show that MCG typically converges for a majority of linear functions representing circuits of 32 lines or less. A linear function, introduced in [1], for which MCG converges is shown in Fig. 7. Fig. 8 illustrates MCG, AECM, and "Algorithm 1" syntheses of this linear function. The total CNOT gate counts are 12 for MCG, 13 for AECM, and 15 for "Algorithm 1".

$$\begin{bmatrix} 1 & 1 & 0 & 0 & 0 & 0 \\ 1 & 0 & 0 & 1 & 1 & 0 \\ 0 & 1 & 0 & 0 & 1 & 0 \\ 1 & 1 & 1 & 1 & 1 & 1 \\ 1 & 1 & 0 & 1 & 1 & 1 \\ 0 & 0 & 1 & 1 & 1 & 0 \end{bmatrix}$$

**Fig. 7.** A 6×6 linear function from [1].

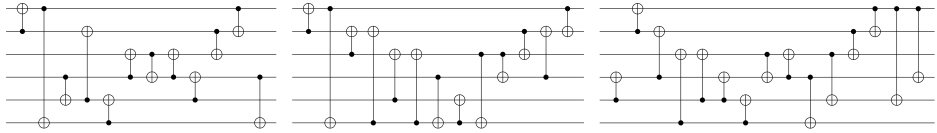

**Fig. 8.** MCG (left), AECM (middle), and "Algorithm 1" (right) syntheses of the linear function from Fig. 7.

Fig. 9 shows MCG synthesis of the linear function in Fig. 7 using Equation 2 which is an alternative cost function based on the sparseness of a linear function and its inverse. This creates a synthesis method that incrementally approaches a low-cost permutation of the identity matrix as it searches for efficient two-CNOT-gate functions. An approach of synthesizing a permutation of a linear function is useful when the output line order is flexible or when the cost of a SWAP gate is negligible in comparison to the cost a CNOT gate. The resulting line-reordered MCG circuit shown in

Figure 9 employs only eight CNOT gates. The circuit realization of the linear function in Figure 10 can be described as the output vector $y_{Permutation} = [y_2, y_1, y_4, y_6, y_3, y_5]^T$.

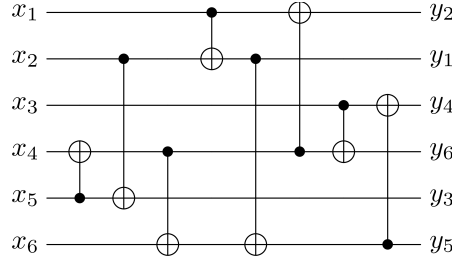

**Fig. 9.** Line-reordering MCG synthesis of the linear function from Fig. 7.

$$\begin{bmatrix} 1 & 0 & 0 & 1 & 1 & 0 \\ 1 & 1 & 0 & 0 & 0 & 0 \\ 1 & 1 & 1 & 1 & 1 & 1 \\ 0 & 0 & 1 & 1 & 1 & 0 \\ 0 & 1 & 0 & 0 & 1 & 0 \\ 1 & 1 & 0 & 1 & 1 & 1 \end{bmatrix}$$

**Fig. 10.** Resulting linear function from line-reordering MCG synthesis which is a permutation of the linear function from Fig. 7.

There are some complications in using line-reordering MCG not described in the above MCG algorithm. The last operation in the MCG algorithm transfers CNOT gates in the output-side CNOT gate list to the input-side CNOT gate list. In the line-reordering MCG variant this transfer must take into account the final permutation remainder function. Therefore for each $M_{output}$ CNOT gate matrix an $M_{input}$ CNOT gate matrix is computed as $M_{input} = P \cdot M_{output} \cdot P^{-1}$. Permutation matrices must be stored with CNOT gate lists for verification and circuit integration purposes. Verification of linear reversible circuit synthesis is usually a straightforward task of applying a CNOT gate list to the identity matrix and testing the result and the input linear reversible function specification for equivalency. In line-reordering MCG verification, each CNOT gate list must apply the associated permutation to the input linear function before testing.

## 4   Tests

The first set of tests was performed for circuits with 8 to 64 lines, and these results are shown in Table I. For each dimension of lines, 100 randomized linear reversible circuits were synthesized with multiple methods. The *n*-wire circuit randomization function used $2n^2$ operations on the identity matrix, and each of these operations represented either a random distant CNOT gate or a random distant SWAP gate. The

MCG method became increasingly slow as the number of lines increased, so MCG testing was stopped at 40 lines. The average CNOT gate count results showed that MCG tended to outperform AECM in functions of up to 24 lines, though which method was best was data dependent. AECM tended to outperform MCG from 28 through 60 lines. At 64 lines "Algorithm 1" tended to outperform AECM, though which method was best was data dependent.

**Table 1.** Comparisons of Linear Reversible Circuit Synthesis Methods (Average Adjacent CNOT Gate Counts).

| Lines | AECM | Algorithm 1 | MCG | MCG Nonconvergent functions |
|---|---|---|---|---|
| 8 | 20.06 | 27.97 | 19.32 | 0 |
| 12 | 43.25 | 62.41 | 40.65 | 0 |
| 16 | 74.06 | 108.1 | 70.94 | 0 |
| 20 | 114.95 | 165.63 | 109.82 | 1 |
| 24 | 167.41 | 233.96 | 161.49 | 0 |
| 28 | 230.59 | 315.74 | 230.68 | 0 |
| 32 | 304.57 | 376.62 | 321.48 | 4 |
| 36 | 393.68 | 468.01 | 418.17 | 42 |
| 40 | 492.84 | 570.12 | 510.63 | 88 |
| 44 | 606.18 | 681.32 | | |
| 48 | 735.64 | 800.09 | | |
| 52 | 873.87 | 930.48 | | |
| 56 | 1028.95 | 1068.58 | | |
| 60 | 1200.66 | 1218.2 | | |
| 64 | 1384.04 | 1373.59 | | |

The second set of tests compared the three synthesis methods with exact minimum syntheses of all 9999360 linear functions of size 5×5 [5]. Table 2 shows the frequency distribution of total CNOT gate counts from exact minimum synthesis of all 5×5 linear functions, a majority of which employ either eight or nine CNOT gates. The results showed that MCG achieved the exact minimum gate count 7175807 (71.76%) times, AECM achieved the exact minimum gate count 5886350 (58.9%) times, and "Algorithm 1" achieved the exact minimum gate count 474738 (4.75%) times. MCG failed to converge in 89 (< 0.001%) linear functions, each time producing a gate count above the exact minimum. This provided strong evidence that nonconvergence in MCG is correlated with increased CNOT gate counts.

**Table 2.** Frequency Distribution of Exact Minimum Syntheses of All 5×5 Linear Reversible Functions.

| CNOT Gates | Functions |
|:---:|:---:|
| 0 | 1 |
| 1 | 20 |
| 2 | 260 |
| 3 | 2570 |
| 4 | 19680 |
| 5 | 117860 |
| 6 | 540470 |
| 7 | 1769710 |
| 8 | 3571175 |
| 9 | 3225310 |
| 10 | 736540 |
| 11 | 15740 |
| 12 | 24 |

$$\begin{bmatrix}
1 & 1 & 1 & 1 & 1 & 0 & 0 & 0 & 1 & 1 & 1 & 0 & 1 & 0 & 1 & 1 \\
0 & 1 & 1 & 0 & 1 & 1 & 0 & 1 & 0 & 1 & 1 & 1 & 1 & 0 & 1 & 0 \\
1 & 1 & 1 & 0 & 0 & 0 & 0 & 0 & 1 & 0 & 1 & 0 & 1 & 0 & 1 & 0 \\
0 & 0 & 1 & 0 & 1 & 1 & 0 & 0 & 0 & 1 & 1 & 1 & 1 & 1 & 0 & 1 \\
1 & 0 & 1 & 1 & 0 & 0 & 0 & 0 & 1 & 0 & 1 & 0 & 0 & 0 & 1 & 0 \\
0 & 1 & 0 & 0 & 0 & 1 & 0 & 1 & 0 & 1 & 0 & 1 & 1 & 1 & 1 & 1 \\
0 & 1 & 1 & 0 & 0 & 1 & 0 & 1 & 1 & 1 & 0 & 0 & 0 & 1 & 0 & 1 \\
1 & 1 & 1 & 0 & 0 & 1 & 0 & 0 & 1 & 0 & 0 & 1 & 0 & 1 & 1 & 1 \\
0 & 1 & 0 & 1 & 1 & 0 & 0 & 1 & 0 & 0 & 1 & 0 & 1 & 0 & 0 & 0 \\
1 & 1 & 0 & 1 & 1 & 0 & 0 & 1 & 0 & 0 & 0 & 1 & 0 & 0 & 0 & 0 \\
0 & 0 & 1 & 0 & 0 & 0 & 0 & 0 & 0 & 1 & 0 & 1 & 1 & 0 & 0 & 0 \\
1 & 1 & 1 & 1 & 1 & 1 & 0 & 1 & 1 & 0 & 0 & 0 & 0 & 0 & 0 & 1 \\
0 & 1 & 0 & 0 & 1 & 0 & 0 & 1 & 0 & 1 & 0 & 1 & 0 & 0 & 1 & 0 \\
0 & 0 & 0 & 0 & 0 & 0 & 1 & 0 & 1 & 1 & 0 & 1 & 0 & 0 & 0 & 1 \\
1 & 1 & 0 & 0 & 0 & 1 & 0 & 1 & 0 & 1 & 1 & 1 & 1 & 1 & 1 & 0 \\
1 & 0 & 1 & 0 & 0 & 1 & 0 & 0 & 1 & 0 & 1 & 0 & 1 & 1 & 0 & 0
\end{bmatrix}$$

**Fig. 11.** 16×16 Test Linear Function.

The last set of tests compared 1000 syntheses of the 16×16 linear function shown in Fig. 11 using probabilistic variations of AECM and MCG named AECMP and MCGP respectively. In these synthesis variations whenever two candidate CNOT gate sequences are compared and found to be of equal cost or gain, one CNOT gate sequence is chosen at random and the other discarded. The purpose of this test was to examine the typical range of total CNOT gate counts produced from AECM and MCG and determine the possible benefits from a multiple-pass approach. The results shown in Table 3 indicated that MCGP showed a difference of 18 CNOT gates between the best and worst syntheses, and the typical spread around the median synthesis was just under four CNOT gates. The results for AECMP indicated a difference of 11 CNOT gates between the best and worst syntheses, and the typical spread around the median synthesis was just above three CNOT gates. Comparing the results it appears that MCGP and, to a lesser degree, AECMP both are likely to benefit from a multiple-pass approach, avoiding a potential high total CNOT gate count resulting from a single-pass synthesis.

**Table 3.** CNOT gate count statistics for 1000 Syntheses of the 16×16 Test Linear Function from Fig. 11.

|  | Maximum | Minimum | Average | Median | Standard Deviation |
|---|---|---|---|---|---|
| MCGP | 77 | 59 | 68.49 | 68 | 3.694098488 |
| AECMP | 84 | 73 | 77.89 | 77 | 3.287241944 |

## 5 Additional Strategies

A preprocessing strategy can be employed in some linear functions to speed up synthesis. For instance, functions with linearly separable components can be mapped into multiple smaller matrices, individually synthesized, and then mapped back to the entire circuit. If any of these smaller matrices are permutation matrices they can be quickly optimally synthesized.

A postprocessing strategy can be employed with all linear reversible circuits. Any section of a synthesized linear reversible circuit that uses a maximum of five lines can be mapped to a 5×5 matrix and synthesized with an exact minimum table. AECM lends itself to this kind of optimization because at some point during processing it will have exactly five diagonal matrix elements to process, whereas MCG may partially process all rows and columns before completing a single diagonalization.

## 6 Conclusion

The bidirectional linear reversible circuit synthesis methods AECM and MCG were introduced. The main heuristic function used to represent cost in AECM and MCG was introduced, as was an alternative cost function. The alternative cost function was used to synthesize a permutation of an input linear reversible function specification, thus eliminating the need for subsequent line reordering. Both of these methods outperformed "Algorithm 1" in the majority of synthesis tests for circuits with less than 64 lines. All test results were verified to be accurate. Probabilistic versions of AECM and MCG were introduced and shown to benefit from a multiple-pass approach.

Although use of "Algorithm 1" seems ideal when the goal is to quickly synthesize thousands of large circuits, the test results indicate that other methods such as AECM and MCG are recommended for smaller circuits, especially when given a significant amount of processing time. Future work in this area will be to use elements from the "Method of the Four Russians" $GF(2)$ matrix inversion approach and other search strategies with the cost minimization approaches introduced here.